\documentclass[allclo]{FBSart}
\usepackage{amsfonts}
\usepackage{amssymb}
\usepackage{amsfonts}
\usepackage{amsmath}
\usepackage{amssymb}
\usepackage{amsthm}
\usepackage{epsf}
\usepackage[usenames]{color}
\newcommand{\cd}{\makebox[0.08cm]{$\cdot$}}

     \font\tenbifull=cmmib10 scaled 1200 
     \font\tenbimed=cmmib9
     \font\tenbismall=cmmib7
\mathchardef\bbkappa="7114 \mathchardef\bbgamma="710D
\mathchardef\bbrho="711A \mathchardef\bbsigma="711B
\mathchardef\bbtau="711C \mathchardef\bbvarrho="7125
\mathchardef\bbvarsigma="7126 \mathchardef\bbxi="7118

\usepackage[dvips]{graphicx}

\runningtitle{Solutions of  the Bethe-Salpeter equation in
Minkowski space} \runningauthor{J.~Carbonell and V.A.~Karmanov}
\begin{document}

\title{Solutions of  the Bethe-Salpeter equation in Minkowski
space and applications to electromagnetic form
factors\thanks{"Relativistic Description of Two- and Three-Body
Systems in Nuclear Physics", ECT*, October 19-23, 2009}}
\author{J.\ Carbonell$^{a}$ and V.A.\ Karmanov$^{b}$}
 \institute{$^a$Laboratoire de Physique Subatomique et Cosmologie, CNRS/IN2P3,\\
53 avenue des Martyrs, 38026 Grenoble, France\\
 $^b$Lebedev Physical Institute, Leninsky
Prospekt 53, 119991 Moscow, Russia}



\maketitle
\begin{abstract}
We present a new method for solving the two-body Bethe-Salpeter
equation in Minkowski space. It is based on the Nakanishi integral
representation of the Bethe-Salpeter amplitude and on subsequent
projection of the equation on the light-front plane. The method is
valid for any kernel given by the irreducible Feynman graphs and
for systems of spinless particles or fermions. The Bethe-Salpeter
amplitudes in Minkowski space are obtained.  The
electromagnetic form factors are computed and compared to the
Euclidean results.
\end{abstract}
%
%
%
%
%
%
%
%
%


\section{Introduction}\label{intr}

Bethe-Salpeter  (BS) equation  for a relativistic bound system was
initially formulated in the Minkowski space \cite{{SB_PR84_51}}.
It determines the binding energy and the BS amplitude. However, in
practice, finding the solution in Minkowski space is made
difficult due its singular behaviour. The singularities are
integrable, but the   standard approaches for solving integral
equation fail.

To overcome this difficulty, Wick \cite{WICK_54}  formulated the
BS equation in the Euclidean space, by  rotating the relative
energy in the complex plane  $k_0\to ik_4$. This "Wick rotation"
led to a well defined integral equation which can be solved by
standard methods. Most of practical applications of the BS
equation have been achieved using this technique
\cite{nakanishi_prog_ThPh,nakanishi_prog_ThPh2} and recent
developments make its solution a trivial numerical task
\cite{NT_FBS_96}. Another method -- the variational approach in
the configuration Euclidean space -- was recently developed
in~\cite{efimov}.

The binding energy provided by the Euclidean BS equation is the
same than the Minkowski one. However, the Euclidean BS amplitude
does not allow to calculate some observables, {\it e.g.}
electromagnetic form factors. The integral providing the  form
factors contains singularities which are different from those
appearing in the BS equation and  whose positions depend on the
momentum transfer. Their existence invalidates the Wick rotation
in the form factor integral. In terms of the Euclidean amplitude,
the form factor can be obtained only approximately, in the so
called static approximation. The error rapidly increases with the
momentum transfer. To avoid this problem, the knowledge of the BS
amplitude in Minkowski space is mandatory.

Thus, fifty years after its formulation, finding the BS solutions
in the Minkowski space is still a field of active research.

Some attempts have been recently made to obtain the Minkowski BS
amplitudes. The approach proposed in \cite{KW,KSW} is based on the
Nakanishi integral representation \cite{nak63,nak71}  of the
amplitudes and solutions have been found for the ladder scalar
case \cite{KW,SA_PRD67_2003} as well as, under some simplifying
ansatz, for the fermionic one \cite{sauli}. Another approach
\cite{bbmst} relies on a separable approximation of the kernel
which leads to analytic solutions. Recent applications  to the
$np$ system can be found in \cite{bbpr}.

In the paper \cite{bs1} we have proposed a new method to find the
BS amplitude in Minkowski space and applied it to the system of
two spinless  particles. In \cite{ck_2f} this approach was
generalized to the two-fermion case. Like in the papers
\cite{KW,SA_PRD67_2003,sauli}, our approach is based on Nakanishi
representation of the BS amplitude. The main difference between
our approach and those followed in \cite{KW,SA_PRD67_2003,sauli}
is the fact that, in addition to the Nakanishi representation, we
use the light-front projection. This eliminates the singularities
related to the BS Minko\-w\-s\-ki amplitudes. Our method is valid
for any kernel given by the irreducible Feynman graphs.

In this paper we give brief review of the approach \cite{bs1} to
find the BS amplitude in Minkowski space, of its results and
applications.

The plan of the paper is the following. In order to present the
method more distinctly, we consider first the case of zero total
angular momentum and spinless particles. In sect. \ref{project},
corresponding equation for the Nakanishi weight function is given.
In sect. \ref{le}, it is applied to the ladder kernel and in sect.
\ref{crL} -- to the cross ladder one. In sect. \ref{2f} the method
is generalized to the two-fermion system. Sect. \ref{emff} is
devoted to application to the electromagnetic form factor, where
advantage of the Minkowski space solution manifests itself in full
measure. Sect. \ref{concl} contains concluding remarks.

\section{Equation for the weight function}\label{project}

For a bound state of total momentum $p$ and in case of equal mass
particles, the BS equation reads
\begin{equation}\label{bs}
\Phi(k,p)=\frac{i^2}{\left[(\frac{p}{2}+k)^2-m^2+i\epsilon\right]
\left[(\frac{p}{2}-k)^2-m^2+i\epsilon\right]} \int
\frac{d^4k'}{(2\pi)^4}iK(k,k',p)\Phi(k',p),
\end{equation}
where $\Phi$ is the BS amplitude, $iK$ the interaction kernel, $m$
the mass of the constituents and $k$ their relative momentum. We
will denote by $M=\sqrt{p^2}$ the total mass of the bound state,
and by $B=2m-M$ its binding energy.

Our approach consists of two steps. In the first one,  the BS
amplitude is expressed  via the  Nakanishi integral representation
\cite{nak63,nak71}:

\begin{equation}\label{bsint}
\Phi(k,p)=\int_{-1}^1\mbox{d}z'\int_0^{\infty}\mbox{d}\gamma'
\frac{g(\gamma',z')}{\left[  k^2+p\cdot k\; z' +\frac{1}{4}M^2-m^2
-  \gamma' + i\epsilon\right]^3}.
\end{equation}

Notice that in this representation,  the dependence on the two
scalar arguments $k^2$ and $p\cd k$  of the BS amplitude is made
explicit by the integrand denominator and that the weight
Nakanishi  function $g(\gamma,z)$ is non-singular. By inserting
the amplitude (\ref{bsint}) into the BS equation one finds an
integral equation, still singular, for $g$.

In the second step, we apply to both sides of BS equation  an
integral transform -- light-front projection  \cite{bs1}  -- which
eliminates singularities  of the BS amplitude.  It consists in
replacing $k\to k+\frac{\omega}{\omega \cd p}\,\beta$ where
$\omega$ is a light-cone four-vector $\omega^2=0$, and integrating
over $\beta$ in infinite limits. We obtain in this way,  a
non-singular equation for the non-singular $g(\gamma,z)$. After
solving it and substituting the solution in eq. (\ref{bsint}), the
BS amplitude in Minkowski space can be easily computed.

This leads (see ref. \cite{bs1} for the detail of calculations) to
the following equation for the weight function $g(\gamma,z)$:
\begin{equation} \label{bsnew}
\int_0^{\infty}\frac{g(\gamma',z)d\gamma'}{\Bigl[\gamma'+\gamma
+z^2 m^2+(1-z^2)\kappa^2\Bigr]^2} =
\int_0^{\infty}d\gamma'\int_{-1}^{1}dz'\;V(\gamma,z;\gamma',z')
g(\gamma',z'),
\end{equation}
This is just the eigenvalue equation of our method. It is
equivalent to the initial BS equation (\ref{bs}). The total mass
$M$ of the system appears on both sides of equation (\ref{bsnew})
and is contained in the parameter
$ \kappa^2 = m^2- \frac{1}{4}M^2. $
As calculations \cite{KW,KSW} show,  $g(\gamma,z)$ may be zero in
an interval $0\le \gamma \le \gamma_0$. The exact value where it
differs from zero is determined by the equation (\ref{bsnew})
itself.

The kernel $V$, appearing in the right-hand side of eq.
(\ref{bsnew}), is related to the kernel $iK$ from the BS equation
by
\begin{equation}\label{V}
V(\gamma,z;\gamma',z')= \frac{\omega\cdot
p}{\pi}\int_{-\infty}^{\infty}\frac{-iI(k+\beta \omega,p)d\beta}
{\left[(\frac{p}{2}+k+\beta\omega)^2-m^2+i\epsilon\right]
\left[(\frac{p}{2}-k-\beta\omega)^2-m^2+i\epsilon\right]},
\end{equation}
with
\begin{equation}\label{I}
I(k,p)=\int \frac{d^4k'}{(2\pi)^4}\frac{iK(k,k',p)}
{\left[{k'}^2+p\cdot k' z'-\gamma'-\kappa^2+i\epsilon\right]^3}.
\end{equation}
For simple kernels $K(k,k',p)$ given by a Feynman graph the
integral (\ref{I}) over $k'$ is calculated analytically. The
integral (\ref{V}) over $\beta$ is also calculated analytically
and it is expressed via residues. The singularities in the BS
equation are removed by the analytical integration over $\beta$.
Equation (\ref{bsnew}) is valid for an arbitrary kernel $iK$,
given by a Feynman graph. The particular cases of the ladder
kernel and of the Wick-Cutkosky model
\cite{WICK_54,CUTKOSKY_PR96_54} are detailed in the next section
and for the cross ladder kernel -- in the paper \cite{bs2}. Once
$g(\gamma,z)$ is known, the BS amplitude can be restored by eq.
(\ref{bsint}).

The variables ($\gamma,z$) are related to the standard light front
(LF) variables  as $\gamma=k_{\perp}^2$, $z=1-2x$. The LF wave
function (see its definition in \cite{cdkm}, for instance) can be
easily obtained by \cite{bs1}:
\begin{equation} \label{lfwf3a}
\psi(k_\perp,x)=\frac{1}{\sqrt{4\pi}}\int_0^{\infty}
\frac{x(1-x)g(\gamma',1-2x)d\gamma'}
{\Bigl[\gamma'+k_\perp^2 +m^2-x(1-x)M^2\Bigr]^2}.
\end{equation}

Eq. (\ref{bsnew}) can be transformed, in principle, to the
equation for the LF wave function $\psi(k_\perp,x)$, though this
requires inverting the kernel in the left-hand side of
(\ref{bsnew}). The initial BS equation (\ref{bs}), projected on
the LF plane, can be also approximately transformed to the LF
equation:
\begin{equation}\label{eq1}
\left(\frac{\vec{k}^2_{\perp}+m^2}{x(1-x)}-M^2\right)
\psi(\vec{k}_{\perp},x)=
-\frac{m^2}{2\pi^3}\int\psi(\vec{k}'_{\perp},x')
V_{LF}(\vec{k}'_{\perp},x';\vec{k}_{\perp},x,M^2)
\frac{d^2k'_{\perp}dx'}{2x'(1-x')}
\end{equation}
with the LF kernel $V_{LF}$ given, for ladder exchange,  in ref.
\cite{bs1}.

It is worth noticing that the LF wave function (\ref{lfwf3a}) is
different from the one obtained by solving the ladder LF equation
(\ref{eq1}), as it was done e.g. in ref. \cite{mc_2000}. The
physical reason lies in the fact that the iterations of the ladder
BS kernel (Feynman graphs) and the ladder LF kernel (time-ordered
graphs) generate different intermediate states. The LF kernel and
its iterations contain in the intermediate state only one
exchanged particle, whereas the iterations of the ladder Feynman
kernel contain also, many-body states with increasing number of
exchanged particles (stretched boxes). This leads to a difference
in the binding energies, which is however small \cite{mc_2000}.
Formally, this difference arises because of the approximations
which are made in  deriving
eq. (\ref{eq1}) from (\ref{bs}). However, for a kernel given by a
finite set of irreducible graphs,  both BS  (\ref{bs}) and LF
(\ref{eq1}) equations are already approximate and it is not
evident which of them is more "physical". The physically
transparent interpretation of the LF wave function  makes it often
more attractive.

\section{Spinless particles. Ladder kernel}\label{le}


As illustration, we give here the kernel $V(\gamma,z;\gamma',z')$
of equation (\ref{bsnew}) for the ladder BS kernel, which  reads:
\begin{equation}
\label{ladder}
iK^{(L)}(k,k',p)=\frac{i(-ig)^2}{(k-k')^2-\mu^2+i\epsilon}.
\end{equation}
We substitute it in eq. (\ref{I}), then substitute (\ref{I}) in
(\ref{V}) and calculate the integrals.
The details of these calculations are given in ref. \cite{bs1} and one obtains:
\begin{equation} \label{Kn}
V(\gamma,z;\gamma',z')=\left\{
\begin{array}{ll}
W(\gamma,z;\gamma',z'),&\mbox{if $-1\le z'\le z\le 1$}\\
W(\gamma,-z;\gamma',-  z'),&\mbox{if $-1\le z\le z'\le 1$}
\end{array}\right.
\end{equation}
where $W$ has the form:
\begin{equation}\label{W_WC_MINK}
W(\gamma,z,\gamma',z')=\frac{\alpha
m^2}{2\pi}\;\frac{(1-z)^2}{D_0} \;\int_{0}^{1}\frac{v^2}{D^2}\;dv
\end{equation}
with $\alpha=g^2/(16\pi m^2)$ and
\begin{eqnarray*}
D_0 &=& \gamma + m^2z^2 + (1-z^2)\kappa^2 \cr
D     &=&   v(1-v)  (1-z') \gamma  +  (1-z) [  (1-v) \mu^2 +  v  \gamma']  \cr
 &+&   v   m^2 \;[  (1-v)   (1-z')z^2 + v{z'}^2(1-z) \;]  +   v   \kappa^2 (1-z)(1-z') \;[  1+z-v(z-z') \;]
\end{eqnarray*}

In the particular case we are considering here, the angular
integral over $v$ can be performed analytically and $W$ obtains the simple expression:
\begin{eqnarray}\label{W}
W(\gamma,z;\gamma',z') &=& \frac{\alpha m^2}{2\pi}
 \frac{(1-z)^2}{\gamma+z^2m^2+(1-z^2)\kappa^2}
\\
&\times& \frac{1}{b_2^2(b_+ -b_-)^3} \left[ \frac{(b_+ -b_-)(2b_+
b_- -b_+ -b_-)}{(1-b_+)(1-b_-)} +2b_+ b_- \log \frac{b_+
(1-b_-)}{b_- (1-b_+)}\right]
\nonumber
\end{eqnarray}
\begin{eqnarray*}
b_\pm &=& -\frac{1}{2b_2} \;\left( b_1 \pm
\sqrt{b_1^2-4b_0b_2}\right),\qquad
b_0 \;=\; (1-z)\mu^2,\\
b_1 &=& \gamma+\gamma' - (1-z)\mu^2 - \gamma' z -\gamma z' +
(1-z')\left[z^2m^2+(1-z^2)\kappa^2\right],
 \cr b_2 &=& -\gamma
(1-z') - (z-z') \left[  (1-z)(1-z')\kappa^2 +(z+z'-zz') m^2
\right].
\end{eqnarray*}

One can show \cite{bs1} that in the case $\mu=0$, which
constitutes the original Wick-Cutkosky model
\cite{WICK_54,CUTKOSKY_PR96_54}, the solution has the form
$g(\gamma,z)=\delta(\gamma)\; g(z)$. The $\gamma$-dependence
disappears from the equation which, after that, exactly coincides
with the Wick-Cutkosky equation
\cite{WICK_54,nakanishi_prog_ThPh,CUTKOSKY_PR96_54}.

\begin{table}[ht!]
\begin{center}
\caption{Coupling constant values as a function of the binding
energy for  $\mu=0.15$ and $\mu=0.5$ obtained with
$\gamma_{max}=3$, $N_{\gamma}=12$, $N_z=10$ and $\varepsilon_R=
10^{-6}$.}
\label{tab1}       
\begin{tabular}{ccc}
\hline\noalign{\smallskip}
$B$ & $\alpha(\mu=0.15)$ &  $\alpha(\mu=0.50)$  \\
\noalign{\smallskip}\hline\noalign{\smallskip}
0.01   &   0.5716           & 1.440 \\
0.10   &   1.437            & 2.498 \\
0.20   &   2.100            & 3.251 \\
0.50   &   3.611            & 4.901 \\
1.00   &   5.315            & 6.712 \\
\noalign{\smallskip}\hline
\end{tabular}
\end{center}
\end{table}

Equation (\ref{bsnew}) has been solved by using the method
explained in Appendix A, i.e. by expanding the solution $g$ on a
bicubic cubic spline basis. By keeping $\varepsilon_R=10^{-6}$
fixed and varying the grid parameters to ensure four digits
accuracy, we obtain for $\mu=0.15$ and $\mu=0.5$ and unit
constituent mass ($m=1$) the values displayed  in  table
\ref{tab1}. They correspond to $\gamma_{max}=3$, $N_{\gamma}=12$,
$N_z=10$. With all shown digits, they are in full agreement with
the results we have obtained, similarly to \cite{mc_2000}, by
using the Wick rotation and the method of \cite{NT_FBS_96}.
Increasing $\varepsilon_R$ to $10^{-4}$ changes at most one unit
in the last digit. This demonstrates the validity of our approach.

\begin{figure*}[ht!]
\vspace{.5cm}
\begin{center}
\includegraphics[width=0.47\textwidth,height=0.2\textheight]{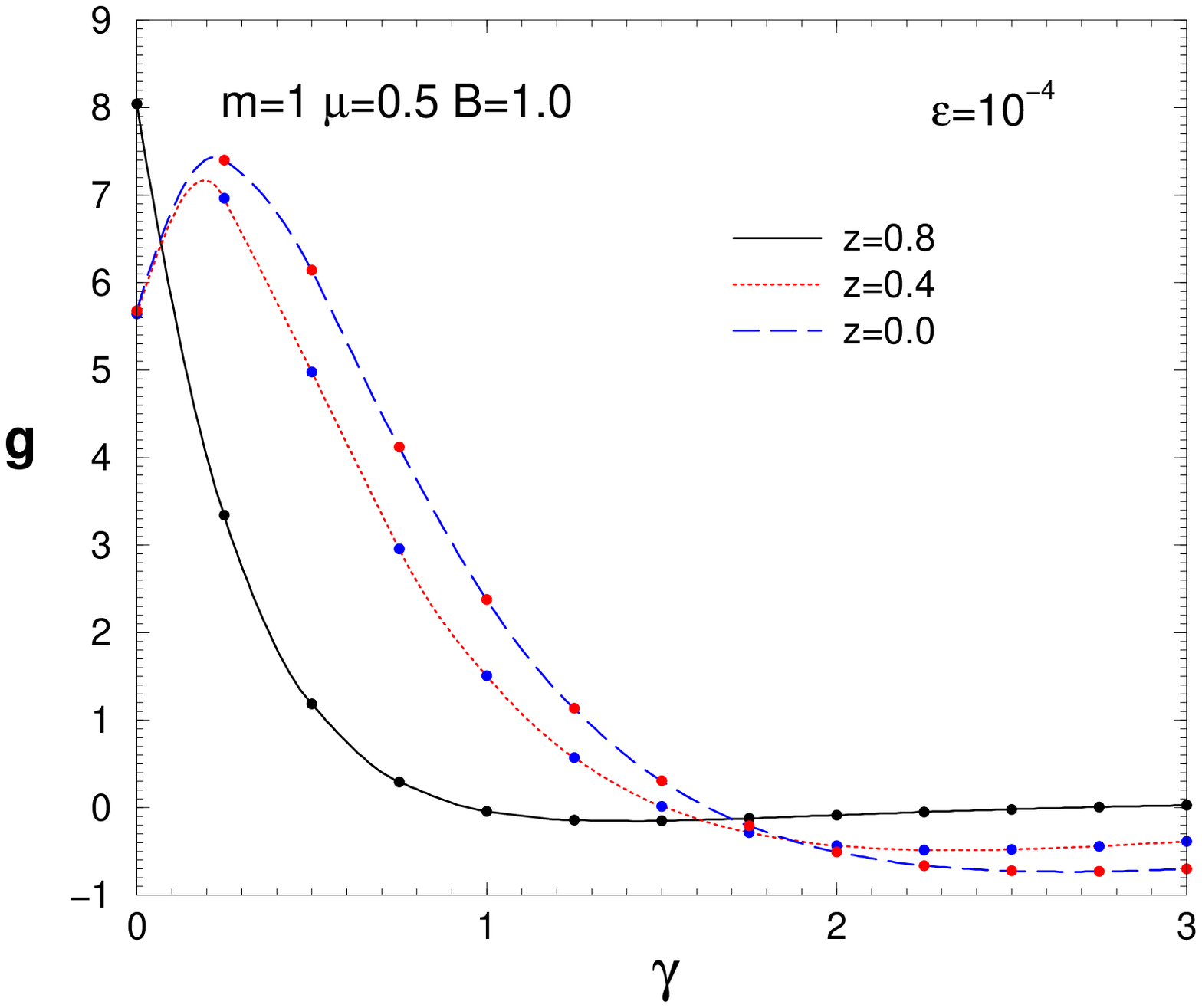}
\hspace{0.5cm}
\includegraphics[width=0.47\textwidth,height=0.2\textheight]{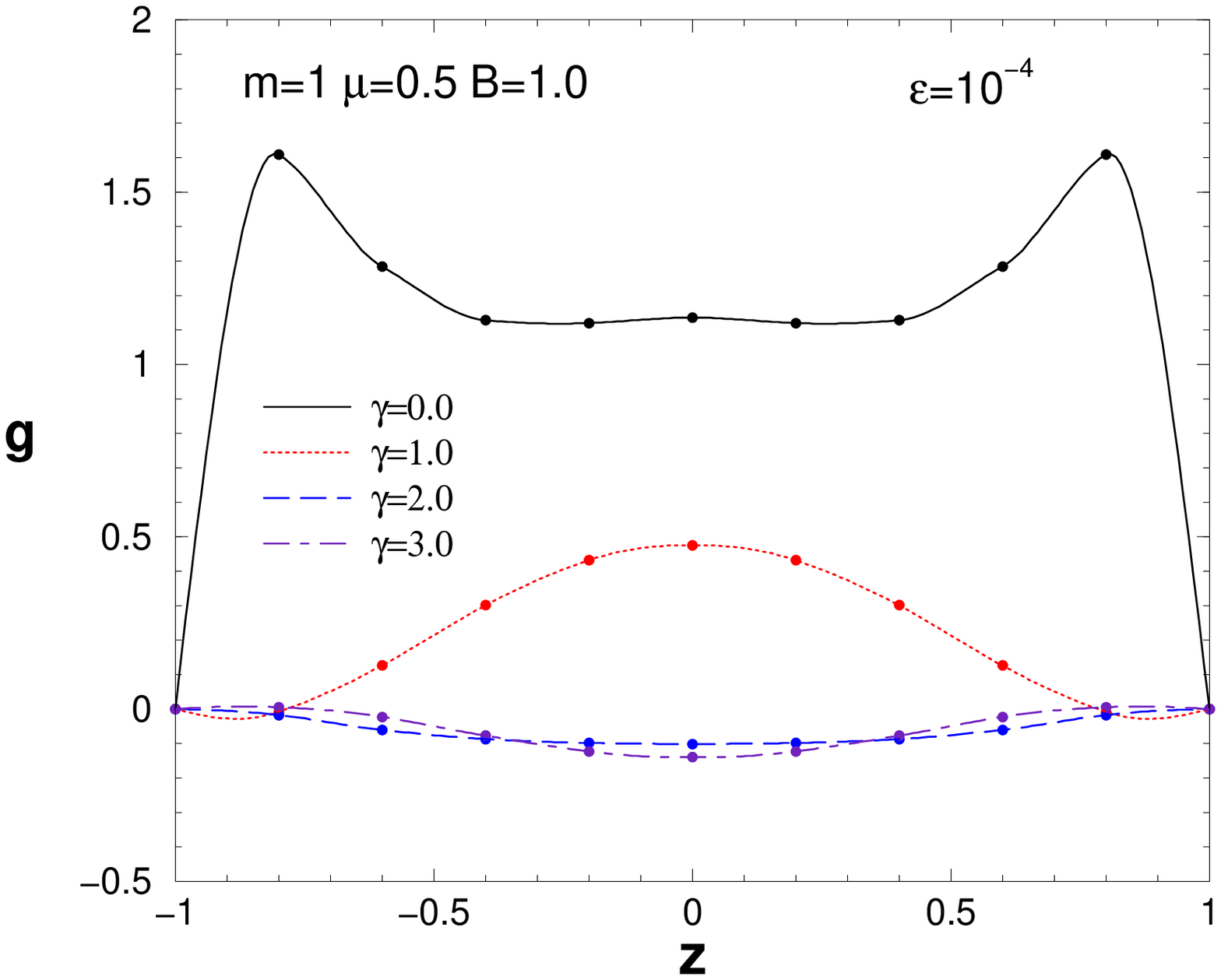}
\vspace{-.5cm} \caption{Nakanishi function $g(\gamma,z)$ for
$\mu=0.5$ and  $B=1.0$. On left -- versus $\gamma$ for fixed
values of $z$ and on right -- versus $z$ for a few fixed values of
$\gamma$.} \label{g_k}
\end{center}
\end{figure*}

The  weight function $g$ for a system with $\mu=0.5$ and  $B=1.0$
is plotted in fig. \ref{g_k}. It has been obtained with
$\varepsilon_R=10^{-4}$ and the same grid parameters than in table
\ref{tab1}. Its $\gamma$-dependence is not monotonous and has a
nodal structure; the $z$-variation  is also non trivial. We have
remarked a strong dependence of $g(\gamma,z)$ relative to values
of the $\varepsilon_R$ parameter smaller than $\sim10^{-4}$, in
contrast to high stability of corresponding eigenvalues.
However the corresponding  BS amplitude $\Phi$ and LF wave
function $\psi$, obtained from $g(\gamma,z)$ by the integrals
(\ref{bsint}) and (\ref{lfwf3a}), show the same strong stability
as the eigenvalues.

\vspace{0.cm}

\begin{figure}[ht!]
\vspace{0.5cm}
 \centering
\includegraphics[width=0.47\textwidth,height=0.2\textheight]{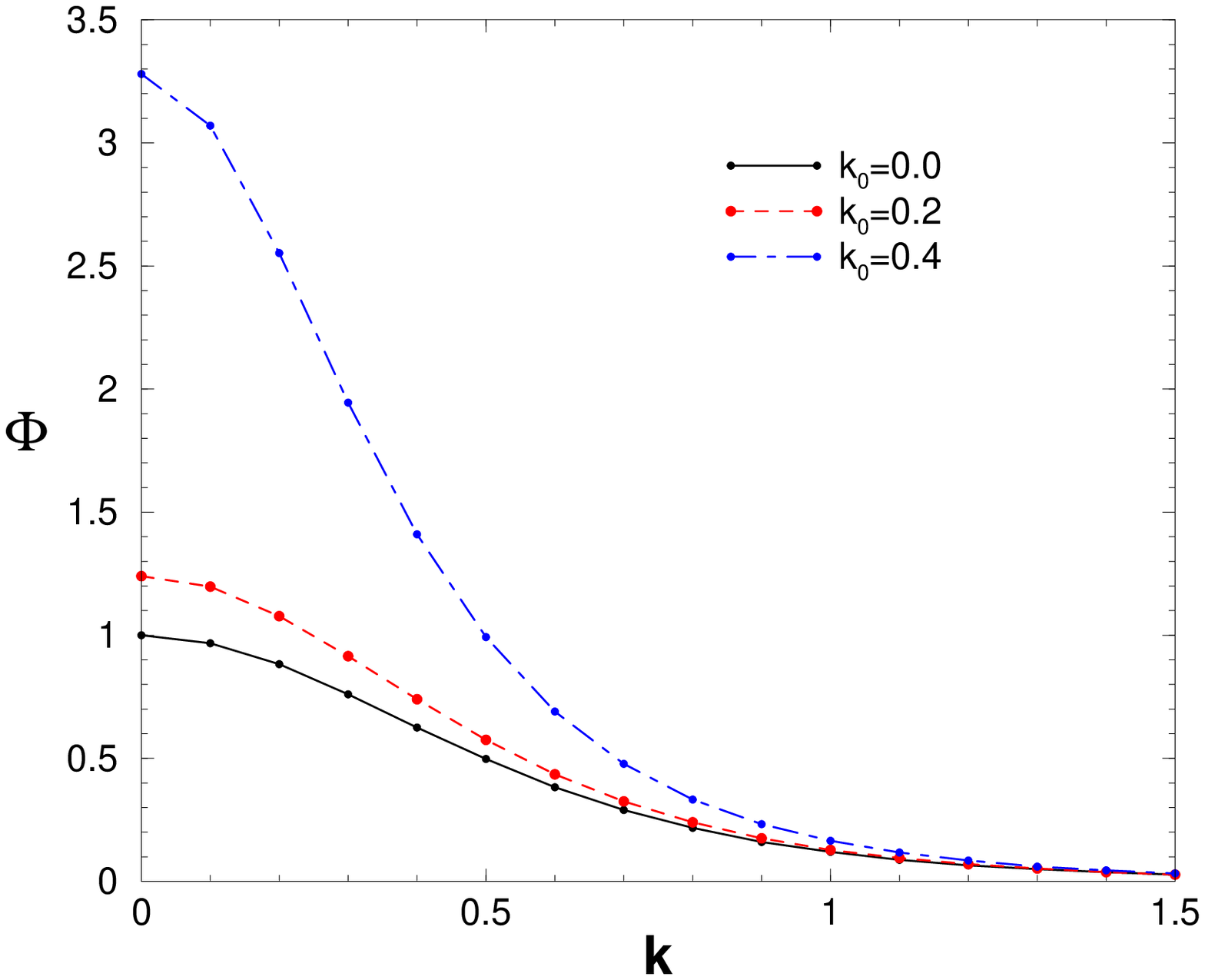}
\hspace{0.5cm}
\includegraphics[width=0.47\textwidth,height=0.2\textheight]{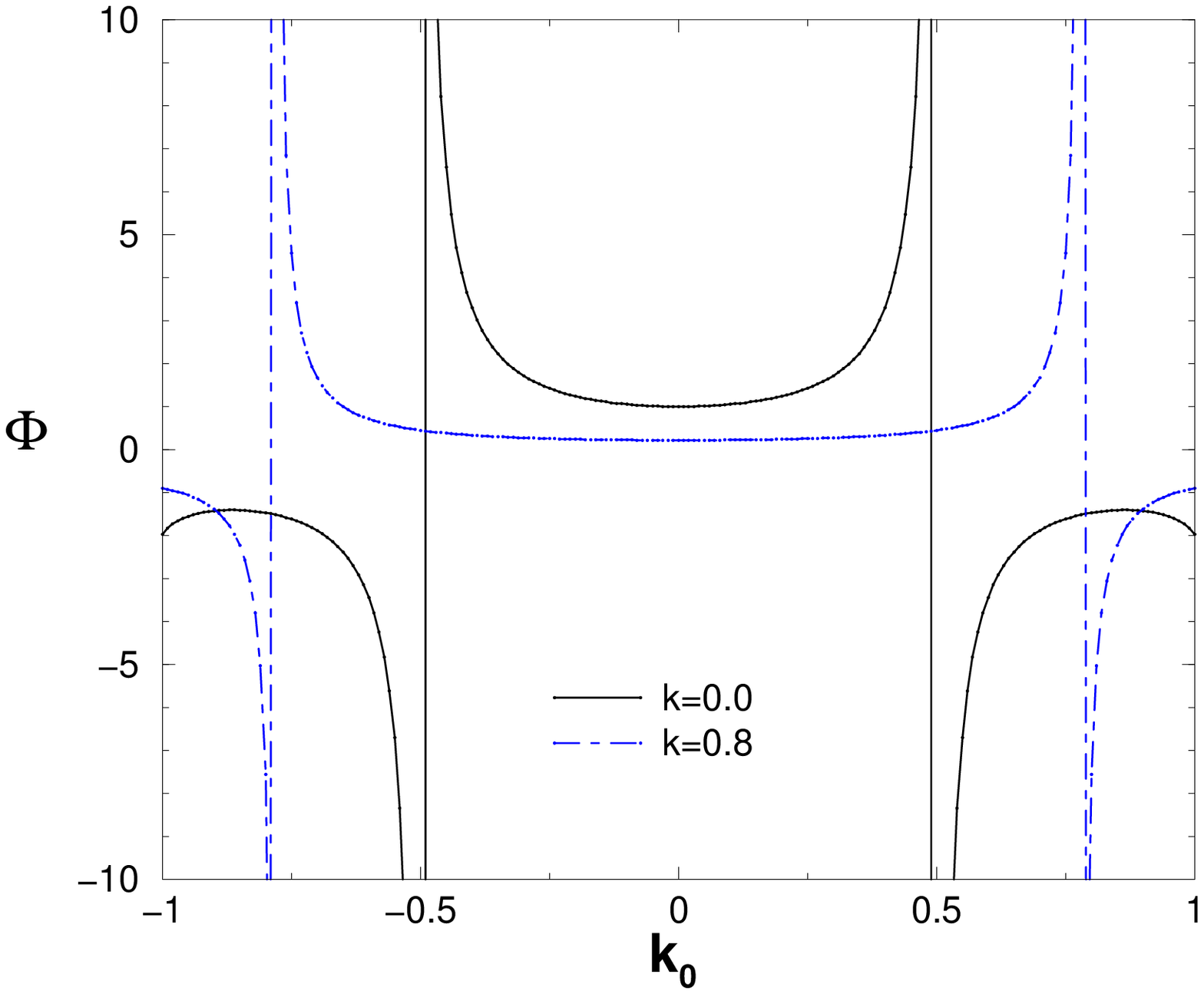}
\vspace{-.5cm}
\caption{BS amplitude  $\Phi(k_0,k)$ for $\mu=0.5$ and  $B=1.0$.
On left versus $k$ for fixed values of $k_0$ and on right versus
$k_0$ for a few fixed values of $k$.} \label{Phi_k}
\end{figure}

The BS amplitude in Minkowski space in the rest frame $\vec{p}=0$
is shown in fig. \ref{Phi_k}. The $k$-dependence is rather smooth
but the $k_0$-dependence, due to poles of the propagators in
(\ref{bs}), exhibits a singular behaviour at $k_0=\pm
\left(\varepsilon_k\pm \frac{M}{2}\right)$, {\it i.e.} moving with
$\vec{k}$ and $M$.

Note that our solution gives also the BS amplitude in Euclidean
space, by substituting in (\ref{bsint}) $k_0=ik_4$. It is
indistinguishable from the one obtained by a direct solution of
the Wick-rotated BS equation \cite{bs1}.

\section{Spinless particles. Cross ladder kernel}\label{crL}


Non-ladder effects, within the same model, using
Feyn\-man-Schwinger representation, were considered in ref.
\cite{NT_PRL_96}. In this work the full set of all irreducible
cross-ladder graphs in a bound state calculation was included.
Refs. \cite{levine,cm2000} contain results on the binding energy
found by solving the BS equation for (L+CL) in Euclidean space. In
the LFD framework, binding energy for (L+SB) kernel is calculated
in \cite{cmp2000}, whereas the (L+CL+SB) contribution is
incorporated in \cite{cm2000}. In \cite{ADT} the effect of the
cross-ladder graphs  in the BS framework  was estimated with the
kernel represented through a dispersion relation. Non-ladder
self-energy  effects have been incorporated in
\cite{Ji,Ji2,SA_PRD67_2003}.

\begin{figure}[ht!]
\begin{center}
\epsfxsize=6cm{\epsffile{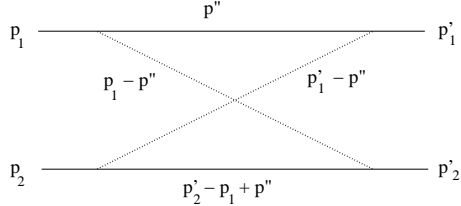}}
\caption{Feynman cross graph.\label{CF}}
\end{center}
\end{figure}

As mentioned, the equation (\ref{bsnew}) is valid for any kernel.
Derivation of $V$ in this equation for the cross-ladder BS kernel,
shown in fig. \ref{CF}, is quite similar but more lengthy, since
the kernel itself is more complicated. We have first to calculate
to the kernel $K^{(CL)}(k,k',p)$ corresponding to the diagram in
fig. \ref{CF}, substitute the result in (\ref{I}), then in
(\ref{V}) and find in this way the cross-ladder contribution to
the kernel $V(\gamma,z;\gamma',z')$ in equation (\ref{bsnew}). The
full kernel -- including ladder and cross-ladder graphs -- will be
written in the form:
 $$
V(\gamma,z;\gamma',z')=V^{(L)}(\gamma,z;\gamma',z')
+V^{(CL)}(\gamma,z;\gamma',z').
 $$
The ladder kernel $V^{(L)}$ is given in refs. \cite{bs1,cdkm}. The
cross-ladder contribution $V^{(CL)}$ was calculated in the paper
\cite{bs2}.

We  compare the results obtained in the BS approach with the
equivalent ones found in Light-Front Dynamics (LFD). The latter
incorporates the graphs with two bosons in flight forming six
cross boxes and two stretched boxes. They were also calculated in
\cite{bs2},

\begin{figure}[h]
\vspace{0.5cm}
\begin{center}
\hspace{0.5cm}
\mbox{\epsfxsize=6.5cm\epsffile{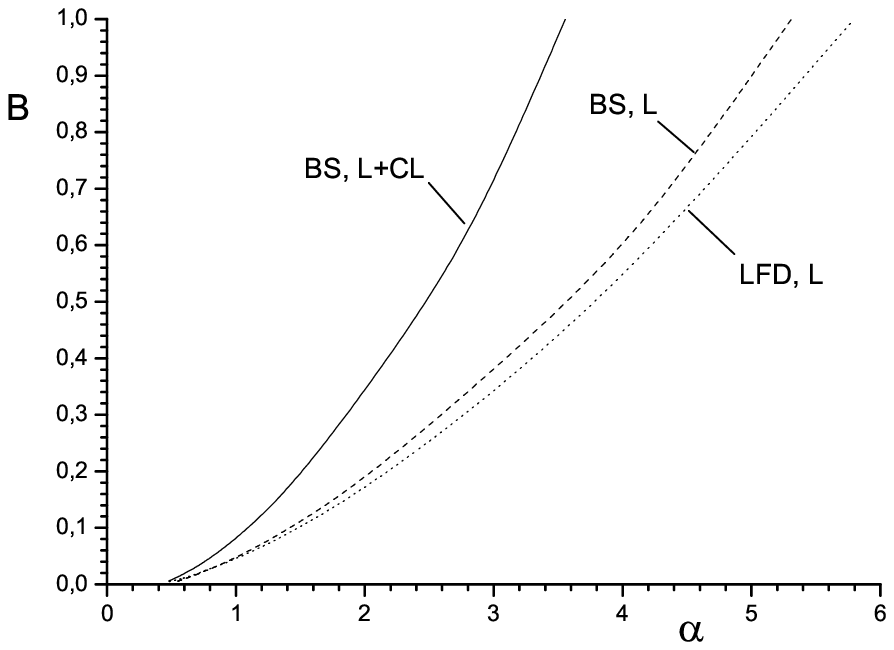}}
\hspace{0.5cm}
\mbox{\epsfxsize=6.5cm\epsffile{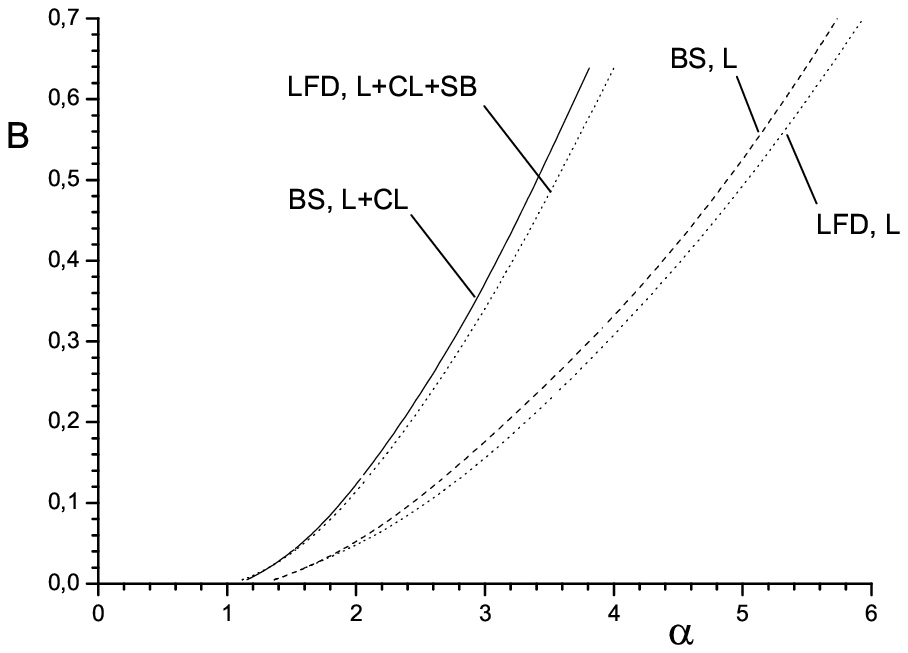}}
\end{center}
\vspace{-0.5cm}
\caption{On left: Binding energy $B$ vs. coupling constant
$\alpha$ for BS and LF equations with the ladder (L) kernels only
and with the ladder +cross-ladder (L+CL) one for exchange mass
$\mu=0.15$. On right: The same as on left but for exchange mass
$\mu=0.5$ and, in addition, binding energy $B$ for LF equation
with the ladder +cross-ladder +stretched box (L+CL+SB) kernel.}
\label{figmu015} \vspace{1.cm}
\end{figure}

The binding energy $B$ as a function of the coupling
constant $\alpha$ is shown in figures \ref{figmu015}
for exchange masses $\mu=0.15$ and $\mu=0.5$ respectively and $m=1$.

We see that for the same kernel -- ladder or (ladder +
cross-ladder) -- and exchange mass  -- $\mu=0.15$ or $\mu=0.5$ --
the binding energies obtained by BS and LFD approaches are very
close to each other. The BS equation is slightly more attractive
than LFD. At the same time, the results for ladder and (ladder
+cross-ladder) kernels considerably differ from each other. The
effect of the cross-ladder is strongly attractive. Though the
stretched box graphs are included, its  contribution to the
binding energy is smaller than 2\% and  attractive as well.

\begin{figure}[h!]
\begin{center}
\includegraphics[width=0.47\textwidth,height=0.2\textheight]{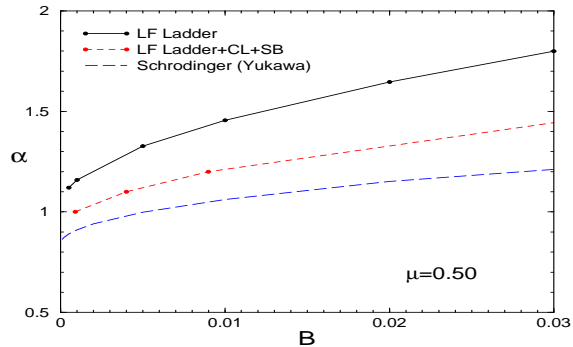}
\end{center}
\vspace{-1.cm}
\caption{Zoom of figure \ref{figmu015} (right) in the zero binding
energy region. The ladder and (ladder +cross ladder +stretched
box) results obtained with the LF equation are compared to the non
relativistic ones (Schrodinger equation with Yukawa
potential).}\label{Zoom}
\end{figure}

The zero binding limit of fig. \ref{figmu015} deserves some
comments. It was found in \cite{mc_2000} that for massive
exchange, the relativistic (BS and LF) ladder results do not
coincide with those provided by the Schr\"odinger equation and the
corresponding non relativistic kernel (Yukawa potential) even at
very small binding energies. Their differences increase with the
exchanged mass $\mu$ and do not vanish in the limit $B\to0$. We
have displayed in fig. \ref{Zoom} a zoom of fig. \ref{figmu015}
(right) for small values of $B$. The cross ladder and stretched
box diagrams reduce the differences but are not enough to cancel it.

We see that the cross-ladder contribution, relative to the ladder
one, results in a strong attractive effect. The BS and LFD
approaches give very close results for any kernel, with BS
equation being always more attractive. These approaches differ
from each other by the stretched-box diagrams with higher numbers
of intermediate mesons. Our results indicate that the higher order
stretched box contributions are small. This agrees with direct
calculations in LFD of stretched box kernel with two-meson states
\cite{sbk} and with calculations of the higher Fock sector
contributions \cite{hk04} in the Wick-Cutkosky model. Calculation
in LFD of binding energy with the stretched box contribution
(L+CL+SB) and its comparison with (L+CL) also shows that the
stretched box contribution is attractive  but small.

The comparison of our results with those obtained in
\cite{NT_PRL_96}, evaluating the binding energy $B_{all}$ for the
complete set of all irreducible diagrams, shows that the effect of
the considered cross ladder graphs, though being very important,
represent only a small part of the total correction. Thus for
$\mu=0.15$ and $\alpha=0.9$ the corresponding binding energies
obtained with BS equation are $B_{L}\approx 0.035$,
$B_{L+CL}\approx 0.06$ and $B_{all}\approx 0.225$.


Qualitatively, our large attractive effect of CL (for $\mu = 0.15$
and $\mu = 0.5$) is in agreement with \cite{levine} (where $\mu =
1$). For $\mu = 0.15$, our numerical results for binding energy,
both for the BS equation in Minkowski space and in the LFD
approach, are close to those found in \cite{cm2000} (within
accuracy of extracting the data from plots). Our results are
smaller than the ones found in \cite{ADT} by a factor 3.

%
\section{Two fermions}\label{2f}


We have considered the following fermion ($\Psi,m$) - meson
($\phi,\mu$) interaction Lagrangians:

({\it i}) Scalar coupling
${\cal L}_{int} ^{(s)}= g\, \bar{\Psi} \phi\, \Psi$
for which $\Gamma_{\alpha}=ig$

({\it ii}) Pseudoscalar coupling
${\cal L}_{int} ^{(ps)}= ig\, \bar{\Psi}\gamma_5 \phi\, \Psi$
for which $\Gamma_{\alpha}=-g\gamma_5$.

 ({\it ii}) Massless vector exchange
${\cal L}_{int} ^{(v)}=  \bar{\Psi} \gamma^{\mu}V^{\mu}\, \Psi$
with  $\Gamma_{\alpha}=ig\gamma^{\mu}$ and
$\Pi_{\mu\nu}=-i{g_{\mu\nu}/q^2}$  as vector propagator.

Each interaction vertex has been regularized with  a vertex form
factor $F(k-k')$ by  the replacement $ g \to gF(k-k') $ and we
have chosen $F$  in the form:
\begin{equation}\label{ffN}
F(q)=\frac{\mu^2-\Lambda^2}{q^2-\Lambda^2+i\epsilon}.
\end{equation}

Let us first consider the case of the scalar  coupling and  the
corresponding  ladder   kernel.
The BS equation for the amplitude $\Phi$ reads:
\begin{equation} \label{bsf1}
\Phi(k,p)=\frac{i(m+\frac{1}{2}\hat{p}+\hat{k})}
{(\frac{1}{2}p+k)^2-m^2+i\epsilon} \left[\int
\frac{\mbox{d}^4k'}{(2\pi)^4}\;\Phi(k',p)
\frac{(-ig^2)\,F^2(k-k')}{(k-k')^2-\mu^2+i\epsilon}\right]
\frac{i(m-\frac{1}{2}\hat{p}+\hat{k})}{(\frac{1}{2}p-k)^2-m^2+i\epsilon}.
\end{equation}

In the case of  $J^{\pi}=0^+$ state, the  BS amplitude has the
following  general form:
\begin{equation}\label{bsf2}
\Phi(k,p)=S_1\phi_1+S_2\phi_2+S_3\phi_3+S_4\phi_4
\end{equation}
where $S_{i}$ are independent spin structures ($4\times 4$
matrices)  and $\phi_{i}$ are scalar functions of $k^2$ and
$p\cdot k$. The choice of $S_i$ is to some extent arbitrary.  To
benefit from useful orthogonality properties we have taken
$$
 S_1= \gamma_5,\quad
 S_2= \frac{1}{M}\hat{p}\,\gamma_5,\quad
 S_3=\frac{k\cdot p}{M^3}\hat{p}\,\gamma_5-\frac{1}{M}\hat{k}\,\gamma_5, \quad
 S_4= \frac{i}{M^2}\sigma_{\mu\nu}p_{\mu}k_{\nu}\,\gamma_5,
$$
where $ \sigma_{\mu\nu}=\frac{i}{2}(\gamma_{\mu}\gamma_{\nu}-
\gamma_{\nu}\gamma_{\mu})$. The antisymmetry of the amplitude
(\ref{bsf2}) with respect to  the permutation $1\leftrightarrow 2$
implies for the scalar functions:
$\phi_{1,2,4}(k,p) = \phi_{1,2,4}(-k,p)$,  $\phi_{3}(k,p)
=-\phi_{3}(-k,p)$.
A decomposition similar to (\ref{bsf2}) was used in \cite{sauli}
to solve the BS equation for a quark-antiquark system but the
solution was approximated keeping only the first term $S_1\phi_1$.

We substitute (\ref{bsf2}) in eq. (\ref{bsf1}), multiply it by
$S_{i}$ and take traces. As we will see, the kernel in the
resulted equation, in contrast to the spinless case, is still
singular. These singularities are integrable numerically. They do
not prevent from finding numerical solution, but they reduce its
precision. This can be avoided by a proper regularization of
equation, multiplying both sides of it by the factor
\begin{equation}\label{eta}
\eta(k,p)=\frac{(m^2-L^2)}{\left[(\frac{p}{2}+k)^2-L^2+i\epsilon\right]}
\frac{(m^2-L^2)}{\left[(\frac{p}{2}-k)^2-L^2+i\epsilon\right]}
\end{equation}
This factor has the form of a product of two scalar propagators
with mass $L$. It plays the role of form factor suppressing the
high off-mass shell values of the constituent four-momenta
$k^2_{1,2}=(\frac{p}{2}\pm k)^2$ and tends to 1  when $L\to
\infty$. In this way, we get the following system of equations for
the invariant functions $\phi_{i}$:
\begin{eqnarray}\label{bsf4p}
\eta(k,p)\;\phi_i(k,p)&=&\frac{\eta(k,p)}
{[(\frac{p}{2}+k)^2-m^2+i\epsilon]
[(\frac{p}{2}-k)^2-m^2+i\epsilon]} \nonumber\\
&\times& \int \frac{\mbox{d}^4k'}{(2\pi)^4} \frac{i
g^2\,F^2(k-k')}{(k-k')^2-\mu^2+i\epsilon}
\sum_{j=1}^4c_{ij}(k,k',p)\phi_j(k',p),
\end{eqnarray}
Since $\eta(k,p)\neq 0$, the equation thus obtained is strictly
equivalent to one where  $\eta(k,p)$ is cancelled. We will see
however that, due to the presence of the $\eta$ factor, the LF
projection modifies the resulting kernels which become less
singular functions.

The coefficients $c_{ij}$ are determined  by traces. We do not
give here their explicit form, which can be found in \cite{ck_2f}.


Then we represent each of the  BS components $\phi_i(k,p)$ by
means of the Nakanishi integral (\ref{bsint})
and, similarly to the scalar case, apply the light-front
projection to the set of coupled equations for the corresponding
weight functions $g_i(\gamma,z)$. As explained in sect.
\ref{project}, this projection, which is an essential ingredient
of our works \cite{bs1,bs2,lc05}, consists in replacing $k\to k+
\frac{\omega}{\omega\cdot p}\beta$ in eq. (\ref{bsf4p}) and
integrating over $\beta$ in all the real domain.

The technical details of the light-front projection are similar to
those  given in ref. \cite{bs1}. We obtain in this way a set of
coupled two-dimensional integral equations:
\begin{equation}\label{eq0f}
\int_0^{\infty}\mbox{d}\gamma'\int_{-1}^1\mbox{d}z' \;
V^g(\gamma,z;\gamma',z')\; g_i(\gamma',z') =
\sum_{j}\int_0^{\infty}\mbox{d}\gamma'\int_{-1}^{1}\mbox{d}z'
\;V^d_{ij}(\gamma,z;\gamma',z') g_j(\gamma',z')
\end{equation}
The kernel $V^g$ and also $V^d_{ij}$ for all types of couplings
and states are given in \cite{ck_2f}. These kernels depend on the
parameter $L$. Closer is $L$ to $m$, smoother is the kernel  and
more stable are the  numerical solutions.  However  the    weight
functions $g_i(\gamma,z)$ as well as  binding energies provided by
(\ref{eq0f})  are independent of $L$. To avoid spurious
singularities in (\ref{eq0f}) due to the $\eta$ factor
(\ref{eta}), $L^2$ must be larger than $\frac{M^2}{2}-m^2$, what
is fulfilled for $L>m$. In practical calculations we have taken
$m=1$ and $L=1.1$.

The kernel  $V_g$  is finite and it vanishes for $z=\pm 1$. For a
fixed values  of $\gamma,z$ and $\gamma'$, $V_g$ is a continuous
function of  $z'$ with a discontinuous derivative at $z'=z$.

\begin{figure}[h!]
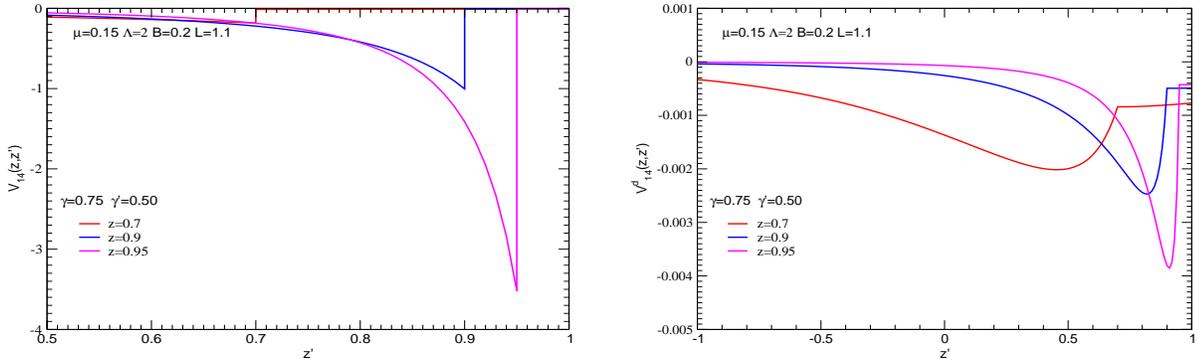

\vspace{.5cm}
\begin{center}
\includegraphics[width=0.47\textwidth,height=0.2\textheight]{V14_2.eps}
\hspace{0.5cm}
\includegraphics[width=0.47\textwidth,height=0.2\textheight]{V14_d_L_1.1_V2.eps}
\caption{Left:  kernel matrix elements
$V_{14}(\gamma,z;\gamma',z')$ for $z=0.7,0.9,0.95$ as a function
of $z'$ and fixed values of $\gamma,\gamma'$. The discontinuity is
finite at a fixed value of $z$ but diverges when $z\to1$. Right:
regularized kernel $V^d_{14}$, plotted  for the same arguments and
$L=1.1m$.} \label{V14}
\end{center}
\end{figure}

As already mentioned, without $\eta$-factor, most of the kernel
matrix elements $V^d_{ij}$ are singular. Namely, they are
discontinuous at $z'=z$. In some cases -- like {\it e.g.} $V_{14}$
displayed in fig. \ref{V14} (left) -- the value of the
discontinuity, although being finite at fixed value of $z$,
diverges when $z\to\pm1$. This creates some numerical difficulties
when computing the solutions $g_i(\gamma,z)$ in the vicinity of
$z=\pm1$. They can be in principle solved by properly taking into
account the particular type of divergence. However the latter may
depend on the particular matrix element, on the type of coupling,
the quantum number of the state and other details of the
calculation.

For $\eta\neq 1$, {\it i.e.}, for a finite value of $L$ in
(\ref{eta}), the $z'$-dependence of the regularized kernels is
much more smooth and therefore better adapted  for obtaining
accurate numerical solutions. In fig. \ref{V14} (right) we plotted
the regularized kernel $V^d_{14}$ as a function of $z'$ for the
same arguments $\gamma,z,\gamma'$ and parameters than in fig.
\ref{V14} (left), where it was calculated without the $\eta(k,p)$
factor. As one can see, the kernel is now a continuous function of
$z'$. A discontinuity of the first derivative however remains at
$z'=z$.

We would like to emphasize again that despite the fact that the
non-regularized and regularized kernels are very different from
each other (compare {\it e.g.}  figs. \ref{V14} left and right)
and that the regularized ones strongly depends on the value of
$L$, they provide -- up to numerical errors -- the same binding
energies and  weight functions $g_i(\gamma,z)$. We construct in
this way a family of equivalent kernels.


\begin{table}[ht!]
\begin{center}
\caption{Left: Coupling constant $g^2$ as a function of  binding
energy $B$   for  the $J=0$ state with scalar (S), pseudoscalar
(PS) and massless vector (positronium) exchange kernels. The
vertex form factor is $\Lambda=2$ and the parameter of the $\eta$ factor $L=1.1$.
%
Right: Coupling constant $g^2$ as a function of  binding energy
$B$ for the positronium $J=0$ state in BS equation in the region
of stability without vertex form factor ($\Lambda\to\infty$), {\it
i.e.} $g<\pi$. They are compared to the non relativistic results.
}\label{tab_B_S_Ps} \vspace{0.3cm}
\begin{tabular}{c|cc |cc |cc |c}
           &  S                 &                         & PS  &     & positronium\\ \hline
$\mu$&  0.15           & 0.50        & 0.15      & 0.50 &0.0
\\ \hline
$B$    & $g^2$         & $g^2$     & $g^2$  & $g^2$  & $g^2$              \\
0.01   &   7.813        &  25.23     &   224.8  &  422.3         &     3.265        \\
0.02   &   10.05        &  29.49     &   232.9  &  430.1         &     4.910 \\
0.03   &   11.95        &  33.01     &   238.5  & 435.8          &     6.263 \\
0.04   &   13.69        &  36.19     &   243.1  & 440.4          &    7.457 \\
0.05   &   15.35        &  39.19     &   247.0  &  444.3         &    8.548 \\
0.10   &   23.12        &  52.82     &   262.1  &   459.9        &  13.15  \\
0.20   &   38.32        &  78.25     &   282.9  &   480.7        &   20.43\\
0.30   &   54.20        & 103.8      &   298.6  &   497.4        &  26.50    \\
0.40   &   71.07        & 130.7      &    311.8 &     515.2      &   31.84 \\
0.50  &   86.95         & 157.4      &    323.1 &     525.9      &   36.62 \\
\end{tabular}
\hspace{2cm} 
\begin{tabular}{c|cc |cc |cc |c}
$B$ &$g^2_{NR}$&$g^2_{BS}$               \\ \hline
0.01   &   2.51          &    3.18        \\
0.02   &   3.55          &    4.65 \\
0.03   &   4.35          &    5.75   \\
0.04   &   5.03          &    6.64    \\
0.05   &   5.62          &    7.38   \\
0.06   &   7.95          &    8.02 \\
0.07   & 11.24          &   8.57 \\
0.08   & 13.77          &   9.06  \\
0.09   & 15.90          &   9.49\\
\end{tabular}
\end{center}
\end{table}

The solutions of eq. (\ref{eq0f}) have been obtained using the  techniques detailed in Appendix A. We have computed the
binding energies and BS   amplitudes, for the $J=0^+$ two fermion
system interacting with  massive  scalar (S) and pseudoscalar (PS)
exchange kernels and for the fermion-antifermion system
interacting with massless vector exchange in Feynman gauge. In the
limit of an infinite vertex form factor parameter
$\Lambda\to\infty$, the later case would correspond to positronium
with an arbitrary value of the coupling constant. All the results
presented in this section are given in the constituent mass units ($m=1$)   and with $L=1.1$.

The binding energies obtained with the form factor parameter
$\Lambda=2$ are given in left table \ref{tab_B_S_Ps}. For the
scalar and pseudoscalar cases,  we present the results for
$\mu=0.15$ and $\mu=0.50$  boson masses. They have been compared
to those obtained in a previous calculation in Euclidean space
\cite{dorkin} using a slightly different form factor, which
differs from our one by a factor.
Once taken into account this correction, our scalar  results are
in  full  agreement (four digits) with \cite{dorkin}. The
pseudoscalar ones show small discrepancies ($\approx 0.5\%$). We
have also computed the binding energies by directly solving the
fermion BS equation the Euclidean space using a method independent
of the one used in \cite{dorkin}. Our Euclidean results are in
full agreement with those given in the table \ref{tab_B_S_Ps}.

\begin{figure}[h!]
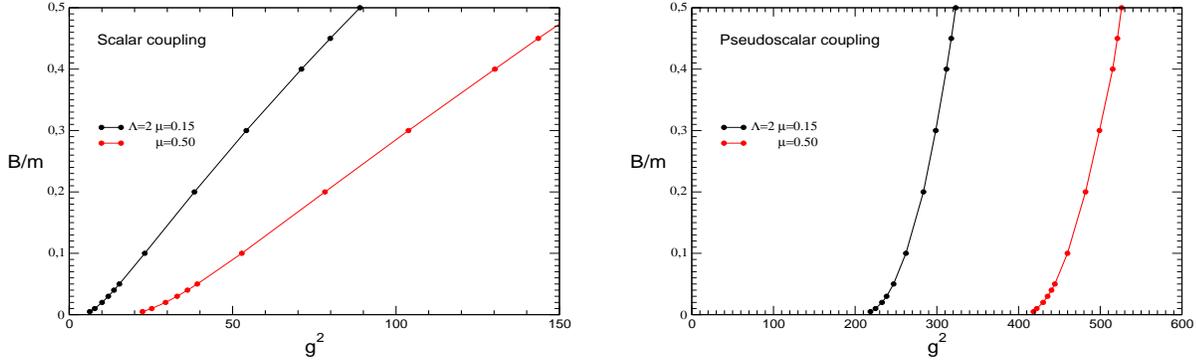

\vspace{.5cm}
\begin{center}
\includegraphics[width=0.47\textwidth,height=0.2\textheight]{B_g2_S.eps}
\hspace{0.5cm}
\includegraphics[width=0.47\textwidth,height=0.2\textheight]{B_g2_Ps.eps}
\caption{Left: Binding energy for scalar exchange v.s. $g^2$ for
$\Lambda=2$,  $L=1.1$, $\mu=0.15$ and $\mu=0.5$.
Right: Binding energy for pseudo scalar exchange v.s. $g^2$ for
$\Lambda=2$, $L=1.1$, $\mu=0.15$ and $\mu=0.5$.}\label{Fig_B_g2_S}
\end{center}
\end{figure}

The $B(g^2)$ dependence for the scalar and pseudoscalar couplings
is plotted in figs. \ref{Fig_B_g2_S}. Notice the different $g^2$
scales of both dependences. The pseudoscalar binding energies are
fast increasing functions of $g^2$ and thus more sensitive to the
accuracy of numerical methods. This sharp behaviour was also
exhibit when solving the corresponding LF  equation
\cite{MCK_PRC68_2003}.

It is worth noticing that the stability properties of the BS
$J^{\pi}=0^{+}$ solutions for the scalar coupling  are very
similar to  the LF ones. In the latter case, we have shown
\cite{MCK_PRD64_2001,MCK_PRD64_2001a}  the existence of a critical
coupling constant $g_c$ below which the system is stable without
vertex form factor while  for $g>g_c$, the system "collapses",
{\it i.e.} the spectrum is unbounded from below. The numerical
value was found to be $\alpha_c=3.72$
\cite{MCK_PRD64_2001,MCK_PRD64_2001a}, which corresponds to
$g_c=6.84$. Performing the same analysis than in our previous work
-- eq. (71) from  \cite{MCK_PRD64_2001,MCK_PRD64_2001a} -- we
found that for BS equation the critical coupling constant is
$g_c=2\pi$, in agreement with \cite{dorkin}. The $10\%$ difference
between the numerical values of $g_c$ is apparently due to the
different contents of the intermediate states in the two
approaches. The ladder BS equation incorporates effectively the
so-called stretch-boxes diagrams which are not taken into account
in the ladder LF results.

The positronium case deserves some comments. First we would like
to notice that in our formalism,  the  singularity of the
Coulomb-like kernels in terms of the momentum transfer
\mbox{$1/(k-k')^2$} is absent. This is a combined consequence of
the Nakanishi transform (\ref{bsint}) -- which allows to integrate
over $k'$ analytically in the right hand side of the BS equation
(\ref{bsf4p}) -- and of the consecutive LF projection integral.
After this integration, the Coulomb singularity  does not anymore
manifest itself in the kernel. This can be explicitly seen in the
kernel of the Wick-Cutkosky model obtained in eq. (22) of our
previous work \cite{bs1}.

A second remark concerns the $\Lambda$ dependence of the
positronium  results. Using the methods developed in
\cite{MCK_PRD64_2001,MCK_PRD64_2001a}  we found that in the BS
approach with ladder kernel there also exists  a critical value of
the coupling constant $g_c=\pi$. Note that, as in the scalar
coupling, the very existence and  the value of this critical
coupling constant is independent on the constituent ($m$) and
exchange $\mu$ masses but depends on the quantum number of the
state.

The ground state  positronium  binding energies without vertex
form factor are given in the left table  \ref{tab_B_S_Ps}  for
values of the coupling below $g_c$, Nonrelativistic results
$g^2_{NR}=8\pi \sqrt{B/m}$ are included for comparison. One can
see that  the relativistic effects  are  repulsive.

\begin{figure}[h!]
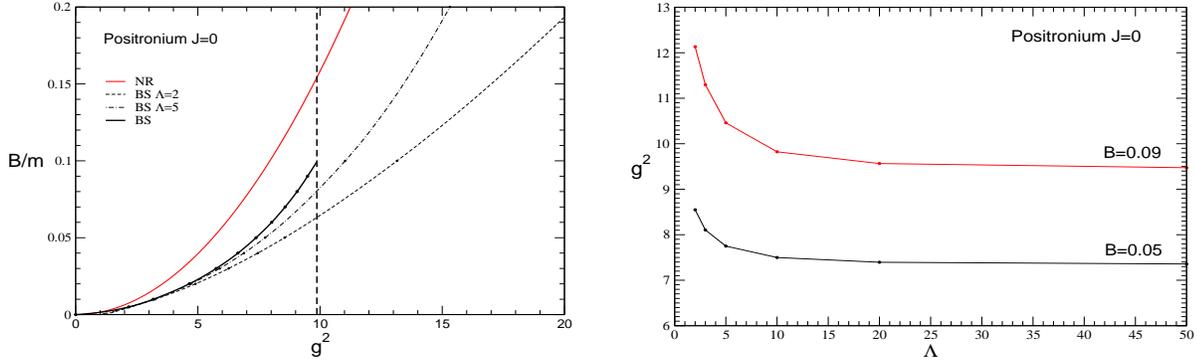

\vspace{.5cm}
\begin{center}
\includegraphics[width=0.47\textwidth,height=0.2\textheight]{B_g2_Positronium.eps}
\hspace{0.5cm}
\includegraphics[width=0.47\textwidth,height=0.2\textheight]{g2_Lambda.eps}
\caption{Left: Binding energy for J=0 positronium state versus
$g^2$ (black solid line) in the stability region $g<g_c=\pi$.
Dashed and dotted-dashed curves correspond to the results for
increasing values of the vertex form factor parameter $\Lambda$.
They are compared to the non relativistic results (red solid
line). Right: $\Lambda$-dependence of $g^2$ for for $J=0$
positronium state for fixed values of binding energies $B=0.1$ and
$B=0.5$.} \label{Fig_B_g2_Positronium}
\end{center}
\end{figure}

These results  are displayed in fig. \ref{Fig_B_g2_Positronium}
(left), black solid line, and compared to the binding energies
obtained with two values of the form factor parameter $\Lambda=2$
(dashed) and $\Lambda=5$ (dot-dashed). The stability region is
limited by a vertical dotted line  at $g=g_c=\pi$. Beyond this
value the binding energy without form factor becomes infinite and
we have found $B(g\to g_c)\approx0.10$.

The inclusion of the form form factor has a  repulsive effect,
{\it i.e.}  for a fixed value of the coupling constant it provides
a binding energy of the system which is smaller than in the
$\Lambda\to\infty$ limit (no cut-off). This is also illustrated in
fig. \ref{Fig_B_g2_Positronium} (right) where we have plotted the
$\Lambda$ dependence of $g^2$ for two different energies. One can
see that the value of the coupling constant to produce a bound
state is a decreasing function of $\Lambda$. The size of the
effect depends strongly on the binding energy but for both
energies the asymptotics is reached at $\Lambda\approx 20$. This
behaviour is understandable in terms of regularizing the short
range singularity of $-1/r^2$ interactions.

\begin{figure}[h!]
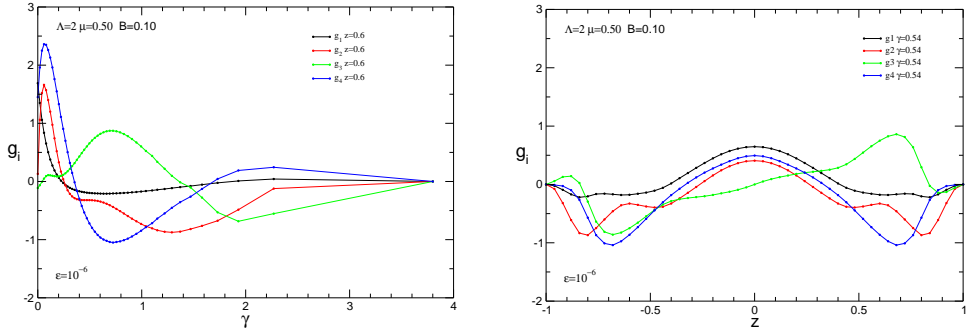

\vspace{0.3cm}
\begin{center}
\mbox{\epsfxsize=6.cm\epsffile{gik_2_0.50_0.10_10M6.eps}}
\hspace{0.5cm}
\mbox{\epsfxsize=6.cm\epsffile{giz_2_0.50_0.10_10M6.eps}}
\vspace{-.5cm}
\caption{Left: Nakanishi weight functions v.s. $\gamma$ for $z=0.6$, for scalar exchange for $\Lambda=2$,  $L=1.1$, $\mu=0.15$
and $\mu=0.5$. Right:  Nakanishi weight functions v.s. $z$ for $\gamma=0.54$.}\label{gig}
\end{center}
\vspace{-.5cm}
\end{figure}

We present in figs. \ref{gig} some examples of the Nakanishi weigh
functions $g_i(\gamma,z)$. They correspond to a $B=0.1$ state with
the scalar coupling  and the same  parameters $\Lambda=2$,
$\mu=0.50$ than in table \ref{tab_B_S_Ps}.  In the left figure is
shown the $\gamma$-dependence for a fixed value of $z$ and in the
right figure the $z$-dependence for a fixed $\gamma$. Notice the
regular behaviour of these functions as well as their well defined
parity with respect to $z$ -- $g_{1,2,4}$ are even and $g_3$ is odd.
As in the scalar case, the $\varepsilon_R$-dependence of $g_i$ is
more important than for the binding energy.

\begin{figure}[h]
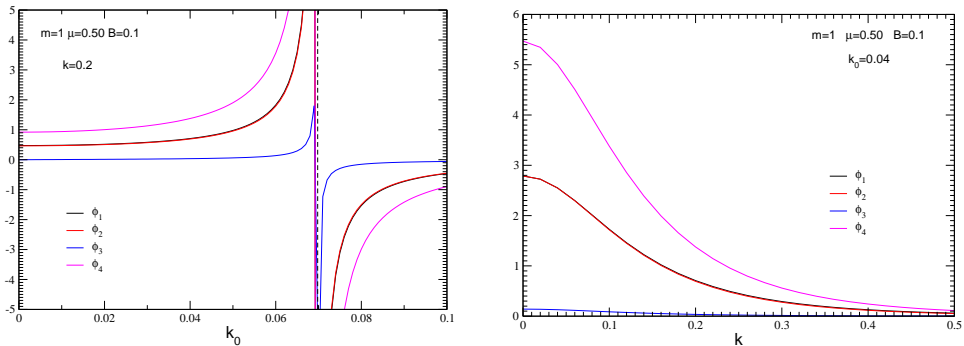

\vspace{.3cm}
\begin{center}
\mbox{\epsfxsize=6.cm\epsffile{Phik0_k_0.2_S_10M2.eps}}
\hspace{0.5cm}
\mbox{\epsfxsize=6.cm\epsffile{Phik_k0_0.04_S_10M2.eps}}
\vspace{-0.5cm} \caption{Bethe-Salpeter Minkowski amplitudes,
corresponding to figs. \ref{gig}, v.s. $k_0$ for $k=|\vec{k}|=0.2$
(on left)   and  v.s. $k=|\vec{k}|$ for $k_0=0.04$ (on right). The
amplitudes $\phi_1$ and $\phi_2$ are
indistinguishable.}\label{Plot_Phi_k0}
\end{center}
\end{figure}

Corresponding BS amplitudes $\phi_i$ are displayed in figs.
\ref{Plot_Phi_k0}. The figure \ref{Plot_Phi_k0} (left) represents
the $k_0$ dependence of $\phi_i$ for a fixed value of
$\mid\vec{k}\mid=0.2$. They exhibit a singular behaviour which
corresponds to the pole of free propagators
$k_0=\epsilon_k-\frac{M}{2}$, like in the scalar case. The figure
\ref{Plot_Phi_k0} (right) represents the $\mid\vec{k}\mid$
dependence of the amplitudes $\phi_i$ for a fixed value
$k_0=0.04$. For this choice of arguments, the amplitudes are
smooth functions of $\mid\vec{k}\mid$, though they will be also
singular for $k_0>\frac{B}{2}=0.05$. Notice that all the
infinities in the BS amplitudes $\phi_i$ come from the free
propagator. However the BS amplitude has other non-analytical
points. For instance it obtains an imaginary part for
$(\frac{p}{2}\pm k)^2>(m+\mu)^2$.
\vspace{-.3cm}

\section{Electromagnetic form factor}\label{emff}
\begin{figure}[h!]
\vspace{0.cm}
\begin{center}
\includegraphics[width=0.4\textwidth,height=0.15\textheight]{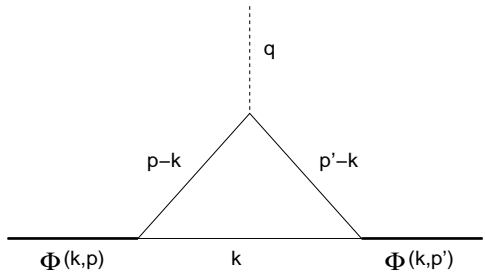}
\vspace{-0.cm}
\caption{E.m. vertex in terms of the BS amplitude.} \label{fffig}
\end{center}
\end{figure}

We demonstrate  in this section the advantages of using the BS
amplitude in Minkowski space for computing the electromagnetic
(em) form factor shown in fig. \ref{fffig}.
For simplicity, we will restrict ourselves to the spinless case for which
one has the following expression in terms of the Minkowski BS amplitude $\Phi_M$
\begin{equation}\label{ffbs}
(p+p')^\mu F_M(Q^2) =-i\int \frac{d^4k}{(2\pi)^4}(p+p'-2k)^\mu \; (m^2-k^2)\Phi_M \left(\frac{p}{2}-k,p\right)\Phi_M  \left(\frac{p'}{2}-k,p'\right)
\end{equation}

As mentioned in the Introduction, the Euclidean BS amplitude
cannot provide the right em  form factor. This fact has been
discussed in some detail in our previous work
\cite{ckm_ejpa,ckm_FBS2008} where two kinds of reasons were
advocated.

\begin{figure}[h!]
\begin{center}
\mbox{\epsfxsize=6.cm\epsffile{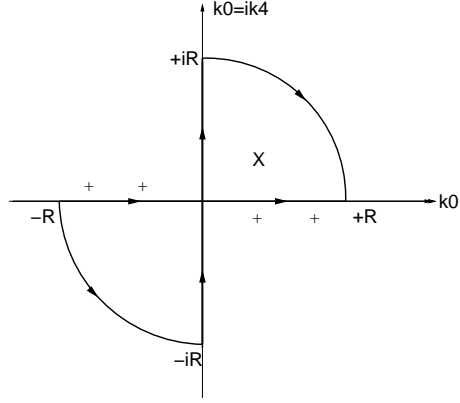}} \caption{Wick rotation
in the  form factor integral.} \label{Wick}
\end{center}
\end{figure}

The  first reason was the impossibility for performing the Wick
rotation in the integral (\ref{ffbs}). Indeed this transformation,
represented schematically in figure \ref{Wick},  consists in
replacing the integral over the  $k_0$ variable along the real
axis  $[-R,+R]$ by  the integral along the imaginary axis
$[-iR,+iR]$. The result is unchanged provided no any singularity
of the integrand is crossed when  rotating anticlockwise the real
into the imaginary integration interval, that is when there are no
singularities in the first or third quadrants. This is the case,
for instance, of those poles associated with the free propagators
(denoted by $+$ in fig. \ref{Wick}) which are located in the
second and fourth quadrants. It turns out however that the
integrand  of (\ref{ffbs}) has singularities in the  $k_0$
variable (denoted by X  in fig. \ref{Wick}) which, for some values
of $\vec{k}^2$, are located in the first or third quadrants. That
prevents from performing a naive Wick rotation, i.e. by simply
replacing  in equation (\ref{ffbs}) $k_0\to ik_4$ and
$\Phi_M\to\Phi_E$, the Euclidean BS amplitudes. Taking properly
into account the contribution of these additional singularities in
the contour integral would require a careful analysis on the
complex plane of the form factor integrand, a task which is hardly
doable even by assuming that $\Phi_M$ and $\Phi_E$ are known
numerically.

This problem is already manifested  at zero momentum transfer in
the simplest model case of a BS amplitude  given by the product of
free  propagators corresponding to the external lines:
\begin{equation}\label{bs0}
\Phi_M=\frac{1}{[k^2-m^2+i\epsilon][(p-k)^2-m^2+i\epsilon]}
\end{equation}
In this case, the integral  (\ref{ffbs}) corresponds to the
Feynman amplitude represented in  fig.  \ref{fffig} with constant
vertices ($g=1$). The form factor  is obtained by multiplying its
result (\ref{ffbs}) by $(p+p')^\mu$. For $p=p'$ it reads:
\begin{equation}\label{ffbs0}
4M^2 F_M(0) =i\int \frac{d^4k}{(2\pi)^4}\frac{(4M^2-4k\cd p)}{[k^2-m^2+i\epsilon][(p-k)^2-m^2+i\epsilon]^2}
\end{equation}
Since the BS amplitude (\ref{bs0}) -- and consequently $F_M$
 -- are not normalized, eq. (\ref{ffbs0}) determines the normalization factor.
Using the Feynman parametrization
$$
\frac{1}{ab^2}=\int_0^1\frac{2xdx}{[(1-x)a+xb]^3}
$$
the integral (\ref{ffbs0}) can be calculated analytically and reads:
\begin{equation}\label{F0}
F_M(0) =\frac{1}{16\pi^2M^3}\left(\frac{4m^2\arctan\frac{M}{\sqrt{4m^2-M^2}}}{\sqrt{4m^2-M^2}}-M\right)
\end{equation}
which for  $m=2$ and $M=3$  takes the numerical value $F_M(0)=4.99241\times 10^{-4}$.

In the rest frame $\vec{p}=0$ the form factor (\ref{ffbs0}) has the form:
\begin{equation}\label{ffbs1}
F_M(0) =\frac{i}{(2\pi)^4 M} \int_0^{\infty}4\pi k^2dk\int_{-\infty}^{\infty} \frac{(M-k_0)dk_0}
{[k_0^2-\vec{k}^2-m^2+i\epsilon][(k_0-M)^2-\vec{k}^2-m^2+i\epsilon]^2}
\end{equation}

In order to check the validity of the "naive Wick rotation" in the
form factor integral (\ref{ffbs1}), we replace there  $k_0=ik_4$
and integrate over $k_4$ in real infinite limits. We obtain in
this way:
\begin{equation}\label{ffbs2}
F_{NW}(0) =\frac{-i}{(2\pi)^4 M} \int_0^{\infty}4\pi k^2dk
\int_{-\infty}^{\infty} \frac{(M-ik_4)idk_4}
{[k_4^2+\vec{k}^2+m^2][(k_4+iM)^2+\vec{k}^2+m^2]^2}
\end{equation}
We have omitted $i\epsilon$ in denominator, since the integrand in
(\ref{ffbs2}) is now non-singular at real $k_4$. For particular
values of the parameters this integral can be computed
numerically. For $m=2$, $M=3$ we get:
$F_{NW}(0)=3.15404\times10^{-4}$ which differs from the value
$F_M(0)$ obtained above, using Minkowski BS amplitudes. This
shows, by a simple example, that performing the Wick rotation to
the variable $k_0$ -- the argument of the BS amplitude -- in the
expression for the em form factor is not allowed.

The reason for this difference is the existence of a pole in the
integrand of (\ref{ffbs1}) created by the zeros of the  the
denominator rightest  factor  which are given by $k_0=
M\mp\sqrt{\vec{k}^2+m^2}\pm i\epsilon$. If $\vec{k}^2<M^2-m^2$,
the  pole corresponding to
$$
k_0=M-\sqrt{\vec{k}^2+m^2}+i\epsilon,
$$
lies  in the first quadrant and it is crossed by the Wick
rotation. The residue at this pole is
$$
{\rm Res}=\frac{-i(M-\sqrt{\vec{k}^2+m^2})}{2^5\pi^2M^3
\sqrt{\vec{k}^2+m^2}(2\sqrt{\vec{k}^2+m^2}-M)^2}
$$
and its contribution to the form factor is given by:
\begin{equation}\label{ffres}
F_{Res}(0) =\int_0^{\sqrt{M^2-m^2}} (2\pi i\, {\rm Res})\; 4\pi k^2 dk
\end{equation}
Numerical calculation for   $m=2$ and $M=3$ gives
$F_{Res}(0)=1.83837\times 10^{-4}$ and the sum of both
contributions is in perfect agreement with the value $F_M(0)$.
$$
F_{NW}(0)+F_{Res}(0)=3.15404\cdot 10^{-4} + 1.83837\cdot 10^{-4}
=4.99241\cdot 10^{-4}
$$

This example illustrates, hopefully clearly, how the singularities
in the first quadrant indeed appear and that they prevent from the
Wick rotation. Taking properly into account their contribution
restores the form factor value. However, this cannot be done if
one knows only the numerical values of the Euclidean BS amplitude.

\bigskip
The second reason results from the fact that even neglecting  the
above explained impossibility of the Wick rotation in the variable
$k_0$, i.e., neglecting the contribution of the singularities denoted by X
in fig. \ref{Wick},  we still cannot compute the form factor
in terms of the Euclidean BS amplitude at rest. Indeed, the
amplitude $\Phi(k,p)=\Phi(k^2,p\cd k)$ depends on the scalars
$k^2=k_0^2-\vec{k}^2$ and $p\cd k=p_0k_0-\vec{p}\cd \vec{k}$.
After Wick rotation $k_0=ik_4$, the first scalar becomes $k^2=-(k_4^2+\vec{k}^2)$.
Suppose that we are working at the rest frame $\vec{p}=0$. Then the
second scalar turns into $p\cd k=p_0k_0=iM k_4$. This purely
imaginary value is just the argument of the Euclidean BS amplitude
and we do not cause any problem.

However, for non-zero momentum transfer $Q^2=-(p-p')^2$ we have
$\vec{p'}\neq 0$. The argument  $p'\cd k$ of the amplitude
$\Phi(k,p')$  -- right vertex in fig \ref{fffig} -- transforms in
the Wick rotation as $p'\cd k=p'_0k_0-\vec{p'}\cd \vec{k}=
ip'_0k_4-\vec{p'}\cd \vec{k}$,   i.e. a  complex value. The
amplitude with complex --  not purely imaginary -- value of its
argument $p'\cd k$ is not the Euclidean one, which enters the left
vertex. Therefore, even after performing the invalid Wick
rotation, the form factor integrand, cannot be expressed via BS
amplitude obtained by real boost from the Euclidean BS amplitude
at rest.

To avoid the complex boost one can solve the Euclidean BS equation
in a moving frame with $\vec{p}\neq0$ and obtain in this way the
BS amplitude for any value of $\vec{p}$. This approach was
followed in \cite{Maris,Maris2} in the framework of the quark
model. The form factor was computed in terms of the Euclidean
amplitudes at $\vec{p}\neq0$ but without taking into account the
contribution of the singularities discussed above.

\bigskip
Another approximate way to compute the form factors,  widely used
in the literature, is the so called static approximation
\cite{zt}. It consists in replacing the  BS amplitude in a moving
frame $\Phi(k;p'_0,\vec{p'})$ by the rest frame amplitude
$\Phi(k;p'_0=M,\vec{p'}=0)$  boosting only  the spatial part of
its argument $k$, i.e. letting unchanged   the time-component
$k_0$. This means in practice that  the left vertex of fig.
\ref{fffig} involves the Euclidean BS amplitude  with $\vec{p}=0$,
$\Phi_E(k_4,\vec{k})$,  while the right vertex involves
$\Phi_E(k_4,\vec{k}+\vec{p'})$. In this way, the form factor is
approximately expressed via Euclidean BS amplitude. The accuracy
of static approximation was estimated in \cite{zt} perturbatively,
at relatively small $Q^2$.

\bigskip
When using the Minkowski BS amplitude, represented via Nakanishi
integral (\ref{bsint}), we can calculate the integral of $d^4k$
analytically and express the form factor in terms of the computed
Nakanishi weight function $g(\gamma,z)$. The procedure is now
theoretically safe and the precision of the result depends only on
the accuracy of the numerical solution of the $g(\gamma,z)$
functions.

\begin{figure}[h]
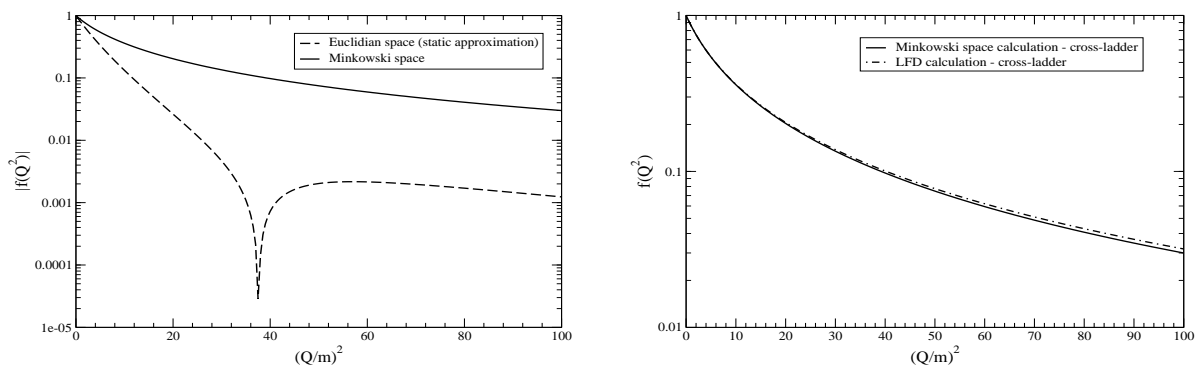

\vspace{0.5cm}
\begin{center}
\includegraphics[
width=0.47\textwidth,height=0.2\textheight]{ff_lcl_mink_estatic_ejp.eps}
\hspace{0.5cm}
\includegraphics[
width=0.47\textwidth,height=0.2\textheight]{{ff_lcl_mink_lfd_ejp.eps}}
\caption{Left: Form factor via Minkowski BS amplitude (solid
curve) and in static approximation (dashed).
\newline
Right: Form factor via Minkowski BS amplitude (solid curve, the
same as at left panel) and in LFD (dot-dashed).} \label{ff2}
\end{center}
\end{figure}

The exact results \cite{ckm_ejpa,ckm_FBS2008}, together with the
static approximation, are shown in fig. \ref{ff2} (left). The
difference between the exact calculation and static approximation
is small at small momentum transfer but it strongly increases with
increase of $Q^2$.

In fig. \ref{ff2} (right) we compare the exact (Minkowski) form
factor with the one found through the LF wave function (\ref{lfwf3a}).
At all $Q^2$ they are almost indistinguishable from each other.

\section{Conclusions}\label{concl}

We presented a new method for obtaining the solutions of the
Bethe-Salpeter equation in  Minkowski space, both for two spinless
bosons and for two fermions. It is based on a Nakanishi integral
representation of the BS amplitude and on subsequent Light-Front
projection.

The binding energies for these systems are calculated. They
coincide with the ones found via Euclidean space solution, thus
providing a validity test of our method. For two fermions, the
solutions for the scalar exchange and positronium states without
vertex form factor ($\Lambda\to\infty$) are 
stable below some critical value $g_c$ of the coupling constant,
respectively $g_c=2\pi$ (scalar exchange) and $g_c=\pi$
(positronium).

The BS amplitudes are obtained in terms of the computed Nakanishi
weight functions. They exhibit a singular behaviour due, on one
hand, to the poles of the free propagators and, on the other hand,
to non analytic branch cuts which are responsible for the
appearance of an imaginary part.

Some applications to em form factors are also presented. We
demonstrate that the Wick rotation in the variable $k_0$ --
argument of the BS amplitude  --  is not allowed. It is prevented
by the singularities which appear in the first quadrant of the
complex $k_0$-plane and whose contributions are hardly evaluable
in practice. Neglecting these contributions, one obtains an
approximate expression which supposes the knowledge of Euclidean
BS amplitude in a non-zero  momentum  frame. The latter requires
to solve numerically a much more involved - though solvable -
equation than at rest. On the contrary, in terms of the BS
amplitude in Minkowski space, represented in the Nakanishi form
(\ref{bsint}), the form factors can be calculated without any
problem.

\bigskip
\appendix
\section{Numerical methods}\label{Ap}
\setcounter{section}{1}

The numerical methods wil be illustrated  by considering  the
general  case of $n_a$ coupled  two-dimensional integral equations
for the Nakanishi weight functions $g_a$ in the form:
\begin{equation}\label{BSMF_mod}
\int_{0}^{\infty} d\gamma'\;\int_{-1}^{+1} dz'\;
V^g_{a}(\gamma, z, \gamma',z') \; g_a(\gamma',z') = \sum_{a,a'=1}^{n_a}\int_{0}^{\infty} d\gamma'\;\int_{-1}^{+1} dz'\;  V^d_{aa'}(\gamma,z,\gamma',z') \; g_{a'}(\gamma',z')
\end{equation}
For the scalar case the number of amplitudes is $n_a=1$ while for
fermions  it depends on the quantum numbers of the state ($n_a$=4
for $J=0$, $n_a=8$ for $ J=1$).

Equation (\ref{BSMF_mod}) has been solved
by expanding the unknown functions $g_a$ on a bicubic spline basis
\cite{PAYNE87} over a compact integration domain $\Omega=
I_{\gamma}\times I_z=[0,\gamma_{max}]\times[-1,+1]$:
\begin{equation}\label{g_spline}
 g_{a}(\gamma,z) =\sum_{i=0}^{2N_{\gamma}+1} \sum_{j=0}^{2N_z+1}
 g_{aij} S_{i}(\gamma)S_{j}(z)   \qquad\forall(\gamma,z)\in \Omega
\end{equation}
The  interval  $I_{x}$ corresponding to  variable $x=\gamma,z$  is divided
into $N_{x}$  subintervals  -- $[x_i,x_{i+1}]$ with  $i=0,\ldots N_x$ --
satisfying   $x_0=x_{min}$ and $x_{N_{x}}=x_{max}$.
The distribution of grid points $G_{x}=\{x_i\}$ are adjusted to
the structure of the solution. By means of expansion
(\ref{g_spline}),  equation  (\ref{BSMF_mod})  can be written in
the form
\begin{equation}\label{BSM5}
 \sum_{a'i'j'} B_{a'i'j'}(\gamma,z) g_{a'i'j'}= \sum_{a'i'j'} A_{a'i'j'}(\gamma,z)g_{a'i'j'}
\end{equation}
with
\begin{eqnarray}
B_{a'i'j'}(\gamma,z)  &=& \int_0^{\infty}  d\gamma' \int_{-1}^{+1} dz' \; V^g_{a}(\gamma,z ,\gamma',z')\;S_{i'}(\gamma') S_{j'}(z')   \;\delta _{aa'}\label{A_ij}\\
A_{a'i'j'}(\gamma,z)  &=& \int_0^{\infty}  d\gamma' \int_{-1}^{+1} dz' \; V^d_{aa'}(\gamma,z ,\gamma',z') \; S_{i'}(\gamma')S_{j'}(z')                                         \label{B_ij}
\end{eqnarray}
The left hand side of (\ref{BSM5}) is in fact diagonal in the number of amplitudes.

The spline functions for the expansion on a variable $x$,  $S_j(x)$,  are defined  once the corresponding grid points $G_x=\{x_0,x_1,\ldots,x_{N_x}\}$ are fixed.
These are $2N_x+1$ functions, usually denoted by $S_0, S_1,\ldots S_{2N_x+1}$.

The functions $S_{2i}$ and $S_{2i+1}$  are associated to the grid point $x_i$ and
have a support limited to the two consecutive intervals surrounding it, i.e.  $[x_{i-1},x_{i+1}]$, a property  that makes easier the computation of integrals (\ref{A_ij}) and (\ref{B_ij}).
Their analytic expressions  are given by
\begin{eqnarray*}
S_{2i}(x)&=&\left\{\begin{array}{ll}
 3\left(\frac{x-x_{i-1}}{x_i-x_{i-1}}\right)^2
-2\left(\frac{x-x_{i-1}}{x_i-x_{i-1}}\right)^3& {\rm if} \;x\in\left[x_{i-1},x_{i}\right]\\
 3\left(\frac{x_{i+1}-x}{x_{i+1}-x_i}\right)^2
-2\left(\frac{x_{i+1}-x}{x_{i+1}-x_i}\right)^3& {\rm if} \;x\in\left[x_{i},x_{i+1}\right]  \\
0  & {\rm if} \;x\notin\left[x_{i-1},x_{i+1}\right]
\end{array}\right.\cr
S_{2i+1}(x)&=&\left\{\begin{array}{ll}
\left[-\left(\frac{x-x_{i-1}}{x_i-x_{i-1}}\right)^2 +
\left(\frac{x-x_{i-1}}{x_i-x_{i-1}}\right)^3\right](x_{i}-x_{i-1})&{\rm if} \;x\in\left[x_{i-1},x_{i}\right]\\
\left[+\left(\frac{x_{i+1}-x}{x_{i+1}-x_i}\right)^2 -
\left(\frac{x_{i+1}-x}{x_{i+1}-x_i}\right)^3\right](x_{i+1}-x_{i})&
{\rm if} \;x\in\left[x_{i},x_{i+1}\right]\cr 0  & {\rm if}
\;x\notin\left[x_{i-1},x_{i+1}\right]
\end{array}\right.
\end{eqnarray*}
and they are such that the values of the solutions at the grid
points are simply given by the even coefficients of the expansion
(\ref{g_spline}), i.e. :
\begin{equation}\label{g_grid}
g_a(\gamma_i,z_j ) = g_{a,2i,2j}
\end{equation}
They are represented in figure 14. 
\begin{figure}[h!]
\begin{center}
\epsfxsize=5cm{\epsffile{spline3.ai}}
\end{center}
\centerline{{\bf \footnotesize Figure 14.} \footnotesize Cubic
spline
functions associated to the grid point $x_i$.}
\end{figure}

Equation (\ref{BSM5}) is validated on a  ensemble
$\{\bar\gamma_{i},\bar{z}_j\}\subset\Omega$ of
$(2N_{\gamma}+2)(2N_z+2)$ "collocation" points which are taken
equal to the N=2 Gauss quadrature abcises inside each subinterval
of $I_{x}$ plus the 2 borders. This leads to a generalized
eigenvalue problem
\begin{equation}\label{Eigen}
\lambda \; B(M) g= A(M)g
\end{equation}
in which matrices $B$ and $A$  represent respectively the integral
operators of the left- and right-hand sides of (\ref{BSMF_mod}) and have the matrix elements
\begin{eqnarray*}
B_{aij ,a'i'j'} &=&  B_{a'i'j'}(  \bar\gamma_i,\bar{z}_j )   \cr
A_{aij ,a'i'j'}&=& A_{a'i'j'}(\ \bar\gamma_i,\bar{z}_j )
\end{eqnarray*}
The unknown coefficients $g_{aij}$ are determined by the solutions
of (\ref{Eigen}) corresponding to $\lambda(M)=1$.

Notice that the value of the solutions at the collocation points are given by the matrix product
\[  g_{a}(\bar\gamma_i,\bar{z}_j) = \sum_{i'j'}  U_{ij,i'j'} \; g_{a'i'j'}  \]
with the spline matrix $U$ given by
\[   U_{ij,i'j'} = S_{i'}(\bar\gamma_i,)S_{j'}(\bar{z}_j) \]

It turns out that the discretized integral operator $B$ has very
small eigenvalues. They are unphysical but  make unstable the
solution of the system (\ref{Eigen}). To regularize $B$, we have
added a small constant $\varepsilon_R$ to its diagonal part
on the form:
\begin{equation} \label{reg}
B\to B +\varepsilon_R \;U
\end{equation}
This procedure allows us to obtain stable eigenvalues with an accuracy of the
same order than $\varepsilon_R$ until values of $\varepsilon_R$ as small as $10^{-12}$.

The property (\ref{g_grid}) is very useful to implement boundary
conditions to the solutions we are interested in. Although not
necessary when working in momentum space  integral equations, it
uses to help the convergence of the numerical algorithms and
allows to avoid unwanted solutions, which can be either due to the
discretisation on a finite box or having other kind of symmetries.
If we know in advance, for instance,  that the solution vanishes
at $z=\pm1$, this implies according to (\ref{g_grid}) to  impose
to the coefficients $g_{a,i,0}\equiv g_{a,i,2N}\equiv0$, that is
in practice to remove them from the expansion (\ref{g_spline})
with the consequent reduction in the dimension of the matrix
equation (\ref{Eigen}).


\end{document}